\documentclass[10pt,conference]{IEEEtran}

\usepackage[T1]{fontenc}
\usepackage[utf8]{inputenc}
\usepackage[scaled=.8]{beramono}
\usepackage{microtype}
\usepackage[all]{nowidow}
\usepackage{balance}
\usepackage{cite}
\usepackage{amsmath,amssymb,amsfonts}
\usepackage{algorithmic}
\usepackage{graphicx}
\usepackage{textcomp}
\usepackage{url}
\usepackage{xspace}
\usepackage{xcolor}
\usepackage{mathrsfs}
\usepackage{todonotes}
\usepackage{pifont}
\usepackage{pdfpages}
\usepackage{booktabs}
\usepackage{longtable}
\usepackage{balance}
\usepackage[abbreviations]{foreign}
\usepackage{cleveref} 

\def\BibTeX{{\rm B\kern-.05em{\sc i\kern-.025em b}\kern-.08em
    T\kern-.1667em\lower.7ex\hbox{E}\kern-.125emX}}

\makeatletter
\newcommand{\linebreakand}{%
  \end{@IEEEauthorhalign}
  \hfill\mbox{}\par
  \mbox{}\hfill\begin{@IEEEauthorhalign}
}
\makeatother

\newboolean{showcomments}
\setboolean{showcomments}{true}
\ifthenelse{\boolean{showcomments}}
{ \newcommand{\mynote}[3]{
		\fbox{\sffamily\scriptsize#1}
		{\small$\blacktriangleright$\emph{\color{#3}{#2}}$\blacktriangleleft$}}}
{ \newcommand{\mynote}[3]{}}

\begin{document}

\title{MLinter:~Learning Coding Practices from Examples---Dream or Reality?}

\author{
\IEEEauthorblockN{
	Corentin Latappy\IEEEauthorrefmark{1}\IEEEauthorrefmark{4},
	Quentin Perez\IEEEauthorrefmark{2},
	Thomas Degueule\IEEEauthorrefmark{1},
	Jean-Rémy Falleri\IEEEauthorrefmark{1}\IEEEauthorrefmark{3},\\
	Christelle Urtado\IEEEauthorrefmark{2},
	Sylvain Vauttier\IEEEauthorrefmark{2},
	Xavier Blanc\IEEEauthorrefmark{1},
	Cédric Teyton\IEEEauthorrefmark{4}
}
\IEEEauthorblockA{\IEEEauthorrefmark{1}Univ. Bordeaux, Bordeaux INP, CNRS, LaBRI, UMR5800, F-33400 Talence, France\\\emph{\texttt{firstname.lastname}@labri.fr}}
\IEEEauthorblockA{\IEEEauthorrefmark{2}EuroMov Digital Health in Motion, Univ. Montpellier \& IMT Mines Ales, Ales, France\\\emph{\texttt{firstname.lastname}@mines-ales.fr}}
\IEEEauthorblockA{\IEEEauthorrefmark{3}IUF, Paris, France}
\IEEEauthorblockA{\IEEEauthorrefmark{4}Promyze, Bordeaux, France}
}

\maketitle

\begin{abstract}
Coding practices are increasingly used by software companies.
Their use promotes consistency, readability, and maintainability, which contribute to software quality.
Coding practices were initially enforced by general-purpose linters, but companies now tend to design and adopt their own company-specific practices.
However, these company-specific practices are often not automated, making it challenging to ensure they are shared and used by developers.
Converting these practices into linter rules is a complex task that requires extensive static analysis and language engineering expertise.

In this paper, we seek to answer the following question: can coding practices be learned automatically from examples manually tagged by developers?
We conduct a feasibility study using CodeBERT, a state-of-the-art machine learning approach, to learn linter rules.
Our results show that, although the resulting classifiers reach high precision and recall scores when evaluated on balanced synthetic datasets, their application on real-world, unbalanced codebases, while maintaining excellent recall, suffers from a severe drop in precision that hinders their usability.
\end{abstract}


\section{Introduction}

Coding practices are essential to software quality.
Some of them are well-known and widely shared by the developer community.
They are even systematically applied in thousands of software projects through the use of linters~\cite{tomasdottir_adoption_2020}, such as ESLint\footnote{https://eslint.org} or checkstyle.\footnote{https://checkstyle.sourceforge.io}
Other coding practices, which are project or company specific, are not supported by linters, even though they are gaining in popularity.
Described in natural language with code examples, they are meant to be applied by developers but always end up buried in inapplicable documentation wikis.

To make specific coding practices more actionable, specialized tools such as CodeQL\footnote{\url{https://codeql.github.com/}} or Semgrep\footnote{\url{https://semgrep.dev/}} provide domain-specific languages to express specific coding practices and engines to apply them automatically.
They, however, require advanced expertise making the definition and application of specific coding practices too complex to be largely used in software projects. 

This lack of support for specific coding practices often hurts the development process since non-compliant code is more likely to land in code review and be discussed repeatedly. This is especially true for the contributions of junior developers who still need to become familiar with project or company practices. 
With Promyze, our industrial partner, we aim to democratize the use of company-specific coding practices. Our dream is to automatically learn practices from examples of compliant and non-compliant source code provided by developers.
The underlying hypothesis is that it would drastically lower the barrier to entry, enabling companies to design and adopt customized coding practices more efficiently.

Machine learning (ML) has yielded promising results for automating coding tasks, such as code completion~\cite{svyatkovskiy_pythia_2019,svyatkovskiy_fast_2021}. However, our use case contrasts with the traditional use of ML because the goal is to learn practices from a small set of examples.
Indeed, in our vision, developers must provide examples of compliant and non-compliant code themselves.
We expect to get, at best, a thousand examples for a given rule, which would even be a relatively high bar requiring a coordinated campaign.
Obtaining a large training dataset is impossible in our setting since we are trying to learn custom practices that are not typically shared by a large community.
For this reason, this work focuses on transfer learning, the state-of-the-art solution to limit the number of examples required.

Therefore, whether ML would be a good solution for our ultimate goal of automating the detection of company-specific coding practices with as few examples as possible remains an open question.
A previous study showed that coding practices could be learned with about 700 examples using decision trees~\cite{ochodek_recognizing_2020}.
However, this is still an upper bound in our industrial context.
Our goal in this paper is to evaluate whether it is possible to train efficient classifiers with even fewer examples.
This paper performs a feasibility study to answer this question.
Our idea is to exploit a popular linter with dozens of coding practices (ESLint) and a vast dataset of open-source projects with conforming and non-conforming code and evaluate how well a state-of-the-art ML technique (CodeBERT~\cite{feng_codebert_2020}) that is compatible with transfer learning can learn the linter's rules with a smaller example budget.

We aim to answer the following research questions:
\begin{itemize}
    \item[RQ1] \textbf{How many examples are required to learn a practice?} Our overarching objective is to be able to learn a practice, ideally using as few examples as possible.
    \item[RQ2] \textbf{What are the best code examples to learn a practice?} One can provide several types of code examples to train a classifier:~examples that do not comply with the practice, examples resulting from enforcing the practice on them (fixes), and examples not related to the practice.
    Is it worth providing fixed code and/or unrelated code to train the classifiers?
\end{itemize}

Our results show that, although the resulting classifiers reach high precision and recall when evaluated on a balanced synthetic dataset, their application on realistic, unbalanced data, while conserving good recall, suffers from a severe precision drop which hinders their usability.

The remainder of this paper is organized as follows.
\Cref{sec:background} introduces some background notions on linters, company-specific coding practices, and our industrial use case.
\Cref{sec:design} presents the overall design of our approach and how we frame the learning task in CodeBERT, \Cref{sec:dataset} details our dataset design process, \Cref{sec:exps} discusses our experimental protocol, and \Cref{sec:results} presents our results.
Finally, \Cref{sec:rw} discusses related work and \Cref{sec:conclusion} concludes.

\section{Background}
\label{sec:background}

\subsection{Coding Practices \& Linters}
\label{subsec:coding_pratices_and_linters}

Linters are static analysis tools that automatically warn developers of possible defects in their code or violations of best practices and coding standards~\cite{beller_analyzing_2016}.
They are routinely used by developers to promote consistency, readability, and maintainability and to improve code reviews.
Linters are popular:~Tómasdóttir et~al.\ find that a quarter of the Javascript repositories they analyze on GitHub use at least one linter, ESLint being the most popular~\cite{tomasdottir_adoption_2020}.

For the sake of illustration, let us detail the \texttt{eqeqeq} coding practice\footnote{\url{https://eslint.org/docs/latest/rules/eqeqeq}} defined in ESLint which states that developers should favor using the equality operators \texttt{===} and \texttt{!==} over the \texttt{==} and \texttt{!=} operators to avoid Javascript's obscure type coercion rules.
A simple example of non-compliant code is \texttt{a~==~b} which should instead be expressed as \texttt{a~===~b}.
The linter's goal is to pinpoint every location in a given codebase that contains non-compliant code for all the coding practices (commonly called rules) it handles.
Violations of the rules are then reported as warnings on the corresponding lines.
Some linters go beyond merely detecting non-compliant code to fix the code and make it comply with the coding practice.
In our example, ESLint can automatically rewrite \texttt{a~==~b} to \texttt{a~===~b}.

Practically speaking, linters typically parse the analyzed codebase to build an abstract syntax tree, visit its nodes, and let rule-specific code hook into the visit steps to implement rule-specific logic.\footnote{\url{https://eslint.org/docs/latest/developer-guide/working-with-rules}}
Developers wishing to implement new rules must acquire extensive knowledge of the tree structure produced by the parser, the linter's internals, and non-trivial static analysis notions.
In the case of company-specific practices, typically designed and used by a limited audience, investing in these development efforts is too costly.

A major concern for the adoption of linters in practice is that they should strive for high precision to minimize the number of false positives and avoid annoying developers.
Indeed, Christakis et~al.\ find that 90\% of developers are willing to accept up to a 5\% false positive rate, and only 24\% of them would tolerate a false positive rate as high as 20\%~\cite{christakis_what_2016}.
Interestingly, they also find that developers favor tools that find fewer bugs over those that generate many false positives.
Tools enforcing coding practices should therefore strive for high precision ($\geq 80\%$) over high recall.

\subsection{Company-specific Coding Practices:~the Promyze Example}

Promyze is a French company specializing in knowledge sharing
for software developers and technical debt management.
Its eponymous tool enables developers to identify and share coding practices within a company or development team.
Using Promyze, developers can identify coding practices directly within their IDE (\eg in Visual Studio) and during code reviews (\eg on GitHub).
Developers are then invited to periodically discuss the identified coding practices during so-called craft workshops, where development teams gather to share knowledge and manually identify positive and negative examples of the practices in their codebases.
Newly-arrived developers are also invited to discovery workshops where they are introduced to existing coding practices to get accustomed to the company's codebase and ease onboarding.

As of November 2022, Promyze's internal catalog hosts 2,825 coding practices created by their customers, out of which 1,933 have at least one associated (positive or negative) example, for a total of 2,830 examples.
Among the 1,828 negative examples, 605 have at least one associated fix.
Besides, Promyze also hosts a public hub of coding practices, open to contributions, which currently hosts 351 practices in a wide range of programming languages stored in 24 catalogs.\footnote{\url{https://bestcodingpractices.dev/}}

An example of a company-specific practice is ``\textit{Avoid \texttt{.toString()}, prefer templating}'',\footnote{\url{https://bestcodingpractices.dev/catalog/632194d9c21bb23fcc68ebb5/631b46fb6ceba90fbd1118f3}} which encourages developers to favor TypeScript's built-in template literal types.
This rule is associated with a positive example and a negative example that have been identified in existing code---in this specific case, the positive example results from enforcing the practice on the negative example:

\begin{minipage}{.8\linewidth}
    \vspace{5pt}
    \ttfamily
    acc[type] = parseInt(acc[type], 10) + 1;
    acc[type] = acc[type].toString();
    \vspace{5pt}
\end{minipage}

\noindent becomes

\begin{minipage}{.9\linewidth}
    \vspace{5pt}
    \ttfamily
    acc[type] = \textasciigrave\$\{parseInt(acc[type], 10) + 1\}\textasciigrave;
    \vspace{8pt}
\end{minipage}

While workshops succeed in identifying coding practices as well as positive and negative examples, they do not scale well for identifying every violation of every practice in huge legacy codebases.
Thus, Promyze's developers experimented with tools such as Semgrep to automate the detection of coding practice violations.
Following hands-on working sessions with their customers, they concluded that their use was too complex for their target audience, as it required advanced expertise.

\section{MLinter}
\label{sec:design}

We now detail our approach to automatically learning coding practices from developer-provided examples to ease their adoption.

\subsection{Overview}
We model our problem of learning coding practices as a binary classification problem, as Ochodek et~al.\ show that it is a successful approach~\cite{ochodek_recognizing_2020}.
Given a coding practice, our goal is to train a binary classifier that takes a line of code as its input and produces  a \textit{compliant} or \textit{non-compliant} label as its output.
This modeling of the problem fits well with our context since we can add new practices by training and deploying a new classifier without affecting the existing classifiers.
Moreover, should the need arise, the classification process can be parallelized to scale up to a larger number of coding practices. 

Linters commonly raise warnings at the level of code lines.
Of course, many coding practices cannot be detected with only one code line as an input.
For instance, the \texttt{indent} practice of \texttt{ESLint}, which ensures that the code is correctly indented, requires contextual information from the surrounding lines.
For the sake of our feasibility study, however, we focus on coding practices that can be identified from a single line of code and leave practices requiring more context to future work.

In our context, two broad categories of code lines can be provided as training examples w.r.t. to a given practice:
\begin{itemize}
    \item \textit{Non-compliant code}:~a source code line that violates the practice;
    \item \textit{Compliant code}:~a source code line that does not violate the practice. A compliant code line can either be found natively in a project---we call it an \textit{extant code line}, or it corresponds to a line that was originally non-compliant and has been manually or automatically fixed to comply with the practice---we call it a \textit{fixed code line}.
\end{itemize}

\subsection{CodeBERT}

Transfer learning is a specific learning method that addresses the problem of insufficient training data. The principle of transfer learning is to first train a machine learning model on source domain data to acquire some knowledge. The model is then retrained on target domain data to specialize it on specific domain knowledge. Several transfer learning models exist. To name a few, we can cite Global Vectors for Word Representation (GloVe)~\cite{pennington_glove_2014}, Word2Vec~\cite{mikolov_distributed_2013} or Bidirectional Encoder Representations from Transformers (BERT)~\cite{devlin_bert_2019}. 

CodeBERT is a large deep learning model based on Transformers~\cite{vaswani_attention_2017} created by Feng \etal~\cite{feng_codebert_2020} and specifically designed for source code. CodeBERT relies on the natural language model BERT. The BERT model outperformed state-of-the-art techniques by a large margin on many Natural Language Processing (NLP) tasks, such as next words/sentence prediction~\cite{devlin_bert_2019}, sentiment analysis~\cite{hoang_aspect-based_2019} and text classification~\cite{garg_bae_2020}. BERT is a pre-trained model: its training is performed on large corpora of natural language documents. Thus, BERT can be used as it stands or adjusted by re-training on a new corpus to create a specific model for a given context or task (transfer learning). Like BERT, CodeBERT is a model allowing the transformation of textual information (lines of code) into vectors and their use for token prediction in lines of code (masked language model) or for classification via re-training.

In our context, we choose CodeBERT for several reasons. First, CodeBERT is a model working purely with raw textual information and does not use any other information such as Abstract Syntax Trees (ASTs) like Code2Seq~\cite{alon_code2seq_2019} or Code2Vec \cite{alon_code2vec_2019} which must transform the input code into ASTs. Thus, we can use this model on single lines of source code, and it relaxes the constraint of parsability on the line. Second, CodeBERT is publicly available through the Hugging Face repository\footnote{\url{https://huggingface.co/microsoft/codebert-base}} and can thus be easily exploited. Third, CodeBERT has been pre-trained on more than 6M lines of code written in six major languages (Go, Java, \textbf{JavaScript}, PHP, Python, and Ruby). This pre-training allows for having less data than the initial training during the specialization (fine-tuning) on our specific classification tasks. The use of six languages in CodeBERT initial training makes the CodeBERT model more generalized than those trained using a single language~\cite{zhou_assessing_2021}. Lastly, CodeBERT has been shown to perform well for many software engineering tasks, for instance, program repair~\cite{mashhadi_applying_2021}, flaky tests prediction~\cite{fatima_flakify_2022}, and defects prediction~\cite{pan_empirical_2021}.

\subsection{Building a Classifier}

\begin{figure}
  \includegraphics[width=\linewidth]{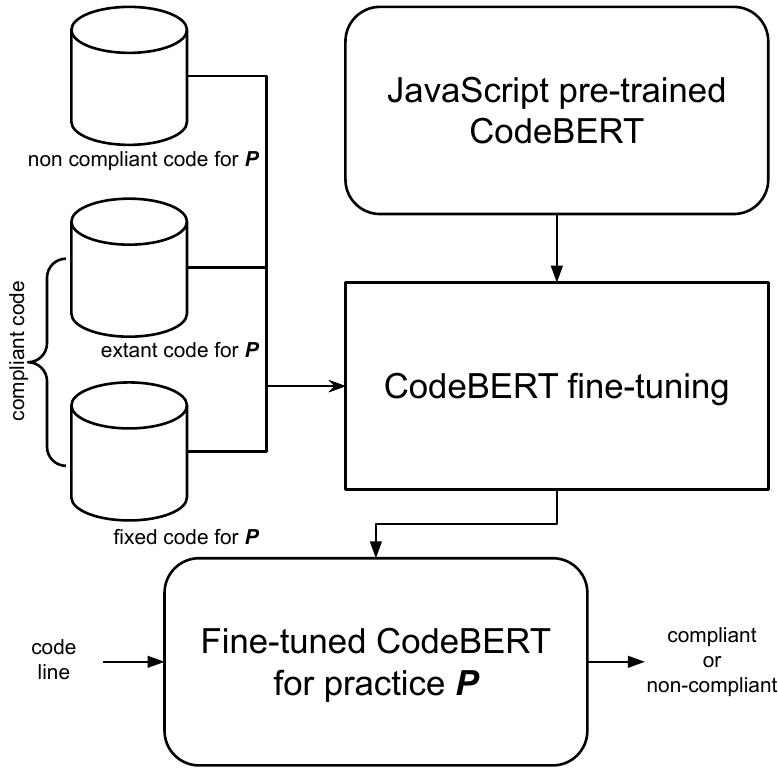}
  \caption{Process to fine-tune CodeBERT for a given practice \textit{P}}
  \label{fig:tuning}
\end{figure}

\Cref{fig:tuning} depicts the process we follow to specialize the pre-trained CodeBERT model to individual coding practices \textit{P}.
This process takes as input the pre-trained CodeBERT model for JavaScript that can be readily downloaded.
It also takes as input a set of training instances of the three previously described types:~\textit{non-compliant code}, \textit{extant code}, and \textit{fixed code} instances.
Finally, the pre-trained model and the training instances are used to produce a new CodeBERT model, fine-tuned for the practice \textit{P}.
This model is then used to classify new source code lines into the \textit{non-compliant} or \textit{compliant} classes.
This process is to be repeated for each of the coding practices that should be learned.
\section{Dataset Design}
\label{sec:dataset}

\begin{figure}
  \includegraphics[width=\linewidth]{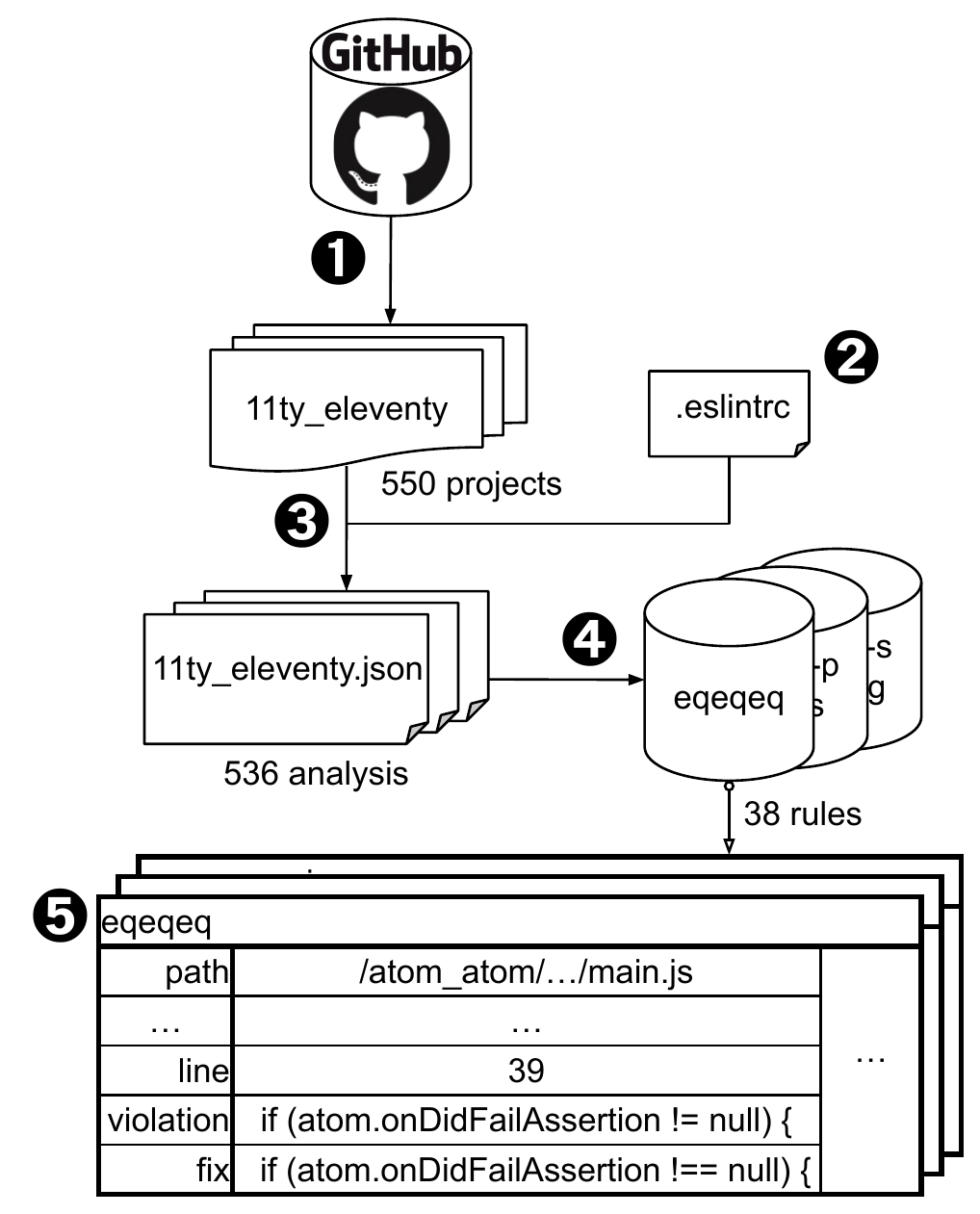}
  \caption{Process to create our dataset}
  \label{fig:dataset_creation}
\end{figure}

Before creating our own dataset, we looked at existing datasets~\cite{ferreira_campos_mining_2019,ochodek_recognizing_2020}.
As explained before, we need our dataset to contain examples of both non-compliant and compliant code for a given coding practice, including compliant code obtained by correcting non-compliant code (\ie \emph{fixed code}).
This was not present in the datasets we analyzed, so we built our own dataset using a linter on popular projects from GitHub.

For this study, we chose to work with JavaScript. Among the linters available for this language, we use ESLint as it is the most popular JavaScript linters~\cite{tomasdottir_adoption_2020}. It is also well documented and has the ability to automatically fix non-compliant code.

We proceeded in three successive steps, detailed below and summarized in \Cref{fig:dataset_creation}:~fetching the JavaScript projects (\eg \texttt{11ty/eleventy} in \Cref{fig:dataset_creation}), configuring ESLint, and extracting and storing results.\footnote{All the scripts used are documented and available for reproducibility:~\url{https://github.com/labri-progress/MLinter}}

\subsection{Project Selection}
\label{subsec:dataset/project-selection}

We use GitHub repositories to create the codebase on which we run ESLint.
Using the GitHub API, we retrieve all public projects whose language is JavaScript and have at least 10,000~stars to obtain a dataset of reasonable size and containing good quality code.
We get a total of 550 repositories that we clone~\ding{202}.
Once all projects have been cloned, we filter all files with the extension \texttt{.min.js}.
Indeed, it is common for JavaScript developers to minify their JavaScript files to speed-up transfer times between servers and clients.
However, it drastically reduces the readability of the code as it usually involves shortening the functions and variables names and removing most white spaces.
Moreover, minified code is not expected to be compliant with the coding practices, as it is only used for deployment.
Therefore, we exclude such code from our training set, as it would introduce noise.

\subsection{Rules Selection}
\label{subsec:dataset/rules-selection}

Before starting the ESLint analysis, we must first select the rules to activate.
Since we need non-compliant code with the corresponding fixed compliant code for each practice, we only consider the fixable rules.
They represent precisely 100 rules.
For this first study, we only focus on rules identifiable on a single code line, as it will be the only data we will provide as an input to the classifier (see~\Cref{sec:design}).
Therefore, by looking at its documentation, we determine for each rule if all the information needed to detect it is on the same line.
Our final ESLint configuration has 54 rules enabled.
Some rules allow us to configure options, resulting in a different analysis behavior.
For simplicity, we let each rule with the default options~\ding{203}.

We modify the build configuration for each project by adding (if not present) a dependency to ESLint with our configuration activated.
Finally, we run ESLint on each project and generate an output in JSON format.
Before running the linter, we have 550 available projects. However, we obtain results for 536 of them (about 2.5\% loss).
For two projects, the analysis failed because of linter errors.
As for the other missing results, ESLint did not find any non-compliant code with the specified configuration.
In this step, we thus obtain 536 JSON files, one per project, containing the ESLint analysis results~\ding{204}.

\subsection{Non-compliant Code and Fixed Code Extraction}
\label{subsec:dataset/violations-fixes-extraction}

For each project, ESLint generates a JSON file containing information about each file of the current project.
Each of these files contains the number of errors found.
For each of these files, we have the number of errors found.
If a file has no errors, we have its filename.
Otherwise, we have the file's complete content and the violations' details.
In almost all cases, each violation is defined by the rule name, details about the file's location, and the associated patch to fix it.
We review all the result files and check two parameters for each error~\ding{205}.
First, related to our need to analyze only one-line rules, we ensure that the violation's start and end lines are the same.
The second check concerns minification.
Despite the filter previously applied on  file names, we observed that minified code was still present in our dataset since not all developers use the \texttt{min.js} convention.
We remove the lines with more than 115 characters to avoid minified code.
To calculate this threshold, we went through 385 random files in our dataset, retrieved the length of all lines, and used the 99\% quantile as our filtering threshold.
We pick only 385 files to avoid to browse all files and save time.
We use Cochran sample size formula with a confidence level at 95\% and a 5\% precision.
When both conditions are met, we record the line number, the violation content, the correction if provided by ESLint, and the path to the file containing the error~\ding{206}. We also record the original GitHub project with the associated commit SHA at cloning time for reproducibility purposes.

Finally, for our protocol needs, we need a minimum of 1,000 examples (violations and fixes) for each rule.
Applying this threshold removes 16 rules, resulting in a final total of 38 rules.

\subsection{Descriptive Statistics}
\label{subsec:dataset/stats}

\begin{figure}
  \includegraphics[width=\linewidth]{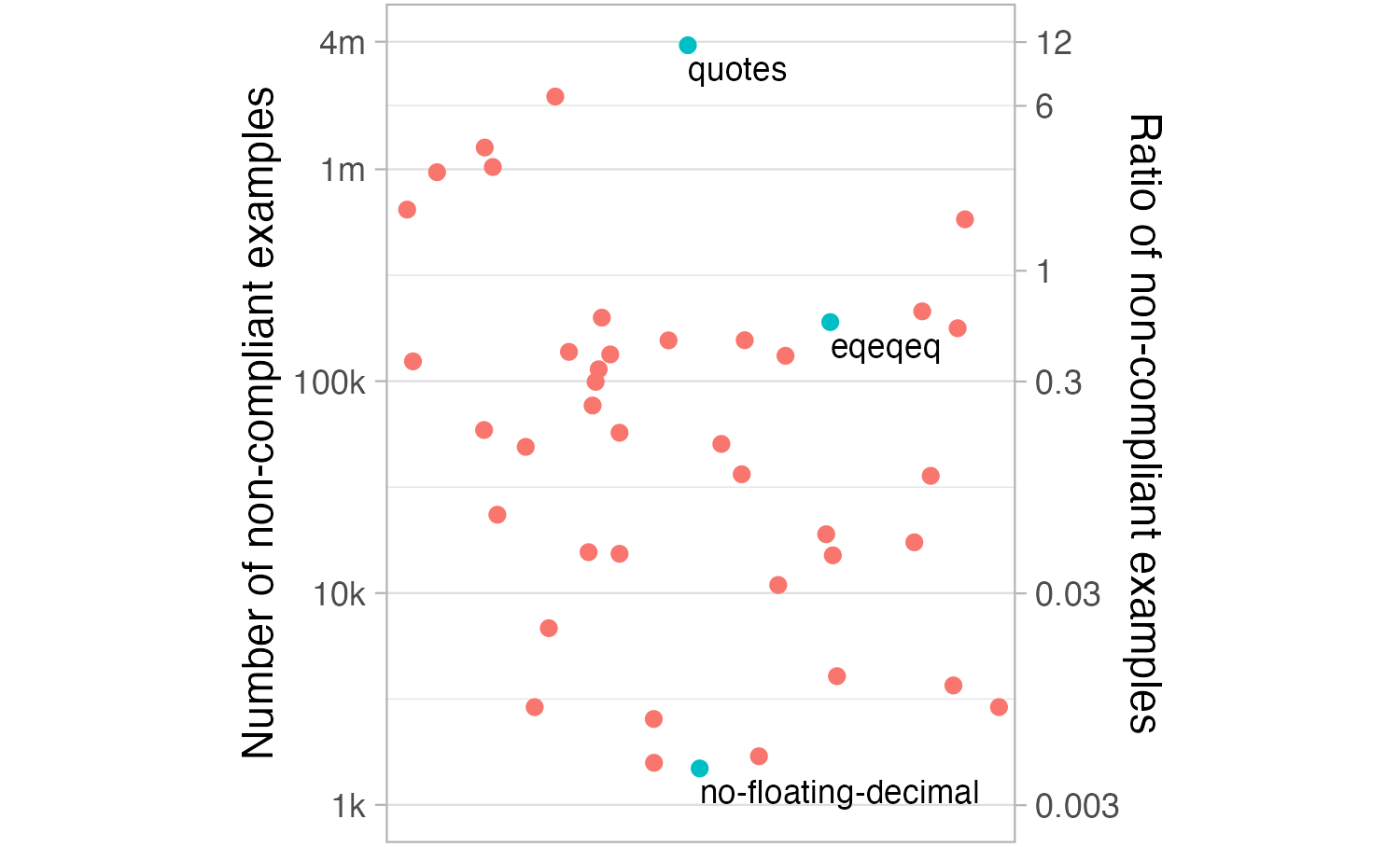}
  \caption{Number of non-compliant examples, with ratio, per rule}
  \label{fig:violation_and_base_rate}
\end{figure}

Our codebase of 550 projects cloned from GitHub contains 218,530 files with more than 33 million lines of code.
Our final dataset contains almost 13 million violations from 38 different rules.
There is an important gap between the most and the least violated rules.
We have almost 4 million examples for the \texttt{quotes} rule and barely 1,480 examples for the \texttt{no-floating-decimal} rule (see~\Cref{fig:violation_and_base_rate}).
An important observation is the ratio between the number of non-compliant examples and the number of lines from our corpus for each rule.
There are 35 rules having a ratio under one percent.
From this observation, we can conclude that most coding practices in our dataset result in extremely imbalanced classification problems~\cite{krawczyk_learning_2016}.
Interestingly, these ratios are consistent with those observed in previous work~\cite{ochodek_recognizing_2020}. Therefore, we conjecture that this distribution of ratios is inherent to the problem of detecting non-compliant code.

\section{Experimental protocol}
\label{sec:exps}

Answering our two research questions requires the construction of several classifiers for different linter rules, with different numbers of examples used to learn a rule and different ratios of compliant and non-compliant code. All of these classifiers are trained on a training set and are validated on a testing set by checking their ability to detect the same rule violations as the linter.

\subsection{Learning Configurations}
\label{sec:configs}

Each classifier targets a coding practice defined by an ESLint rule and checks whether a given line of code complies with the practice.
It is trained on a training set that consists of source code lines, some of which are non-compliant and others compliant.

Regarding the number of lines in the training set, we recall that our main motivation is to check whether a single developer or a small development team can train a classifier by feeding it with a small-enough number of examples. 
Following discussions with our industrial partner, we consider that a set of 10 examples is quite small (easily done by a single developer), 100 is medium (within reach of a development team), and 1000 is large (would require coordinated work from several teams).
We, therefore, consider three sizes (S=10, M=100, L=1,000) for the training sets.

For most practices, we are dealing with extremely imbalanced data where compliant code largely dominates over non-compliant code (see \Cref{sec:dataset}).
To deal with this issue, we use a classical data-level method relying on undersampling compliant code instances to construct balanced training sets with 50\% of instances of compliant and non-compliant code~\cite{krawczyk_learning_2016}.
The 50\% of compliant code in the training sets can either be fixed or extant, as described in \Cref{sec:design}.
Our hypothesis is that the classifier would learn better when presented with some amount of fixed code, which represents a sort of ``border'' between the compliant and non-compliant code instances.
Following this idea, we opt for three training set ratios:~50\% of non-compliant code and 50\% of fixed code (the VF ratio), 50\% of non-compliant code and 50\% of extant code (the VE ratio), and 50\% of non-compliant code, 25\% of fixed code, and 25\% of extant code (the VFE ratio).

We obtain nine learning configurations for each classifier by combining the three sizes (S, M, L) and the three ratios (VF, VE, VFE).
For example, an M/VF classifier is trained on a set of 100 lines composed of 50 non-compliant and 50 fixed code lines.
Note that the special case S/VFE is trained on 10 lines composed of 5 non-compliant code lines, 3 fixed code lines, and 2 extant code lines.

\subsection{Validation Protocol}
\label{sec:validation}

As the number of source code lines we use for our training sets (10, 100, or 1000) is very small compared to the total number of source code lines in our dataset (33M lines), it is likely that the precision and recall obtained by a particular classifier is not representative enough.
To address this threat, we use a validation inspired by the out-of-sample bootstrap validation approach~\cite{efron_estimating_1983} that is the best performing validation approach in~\cite{tantithamthavorn_empirical_2017}.
We train 100 classifiers for each of the nine configurations and 38 linter rules.
The lines included in a given training set are drawn at random with replacement from the whole dataset until the expected size and ratios are obtained.
For instance, for a S/VFE classifier training set, we draw at random 5 non-compliant code lines, 3 fixed code lines, and 2 extant code lines.
In contrast to the classical out-of-sample bootstrap, we do not draw a training set with the same size as the whole dataset because, in our study, we want to assess the effect of the training test size.
This training set is then used to fine-tune CodeBERT as explained in \Cref{sec:design}.
As recommended by Devlin \etal \cite{devlin_bert_2019}, we use the following hyper-parameters for the fine-tuning process: $4$ epoch, a batch size of $16$, a learning rate of $5\mathrm{e}{-5}$ and a eps of $1\mathrm{e}{-8}$.

To validate a classifier, we apply it on a set of source code lines (the testing set), and ask, for each line, whether it is compliant or not.
We then compare the classifier's results with the ground truth established by ESLint on the same lines and calculate its precision, recall, accuracy, and F-score.
To build a testing set, we need to define its size and balance (the percentage of compliant and non-compliant code lines in the set).
We define two approaches for building the testing set: the \textit{balanced} and the \textit{realistic} approaches.
In the balanced approach, we build a testing set with the same size and balance as the training set by drawing instances at random without replacement among the instances not included in the training set to mimic the out-of-sample bootstrap approach.
In contrast to the out-of-sample bootstrap, we do not use all instances not in the training set as the testing set, as our dataset contains millions of code lines, and using it would be computationally too expensive.
In the realistic approach, we build a testing set with a balance similar to real source code files.
We construct a testing set composed of all lines with less than 115 characters contained in 5 source code files drawn at random without replacement among the files that have no common line with the training set while having at least one non-compliant code line w.r.t. the practice.
This testing set aims to approximate the balance of the classes that exist in real code files while having some amount of non-compliant code to compute precision and recall.

In total, our protocol builds 34,200 classifiers (38 rules $\times$ 9 kinds $\times$ 100).
Each classifier learns on its own training set, randomly constructed from a global set of source code lines containing non-compliant, fixed, and extant code lines.
Each classifier is then validated twice according to our two validation approaches.
\section{Results}
\label{sec:results}

To answer our research questions, we now discuss the results obtained by applying the protocol introduced in \Cref{sec:exps} to the dataset designed in \Cref{sec:dataset}.
Specifically, we study the influence of our two main parameters on the performance of the resulting classifiers:~the size of the learning set, and the ratio of non-compliant and compliant lines.

We divide this section along our two validation methods.
We first present the results obtained with the balanced validation in \Cref{subsec:results/balanced}, then the results obtained with the realistic validation in \Cref{subsec:results/real_world}, we discuss the results in \Cref{subsubsec:results/discussion} and finally we conclude with the threats to validity in \Cref{subsec:results/ttv}.

For each classifier, we compute its accuracy, precision, and recall scores.
We aggregate the results obtained for every rule by size and by ratio.
This means that, for a given configuration (\eg M/VF) and validation method (\eg realistic), we aggregate the 3,800 scores obtained by the corresponding classifiers (38 rules $\times$ 100 classifiers).

\subsection{Balanced Validation}
\label{subsec:results/balanced}

\begin{figure}
\includegraphics[width=\linewidth]{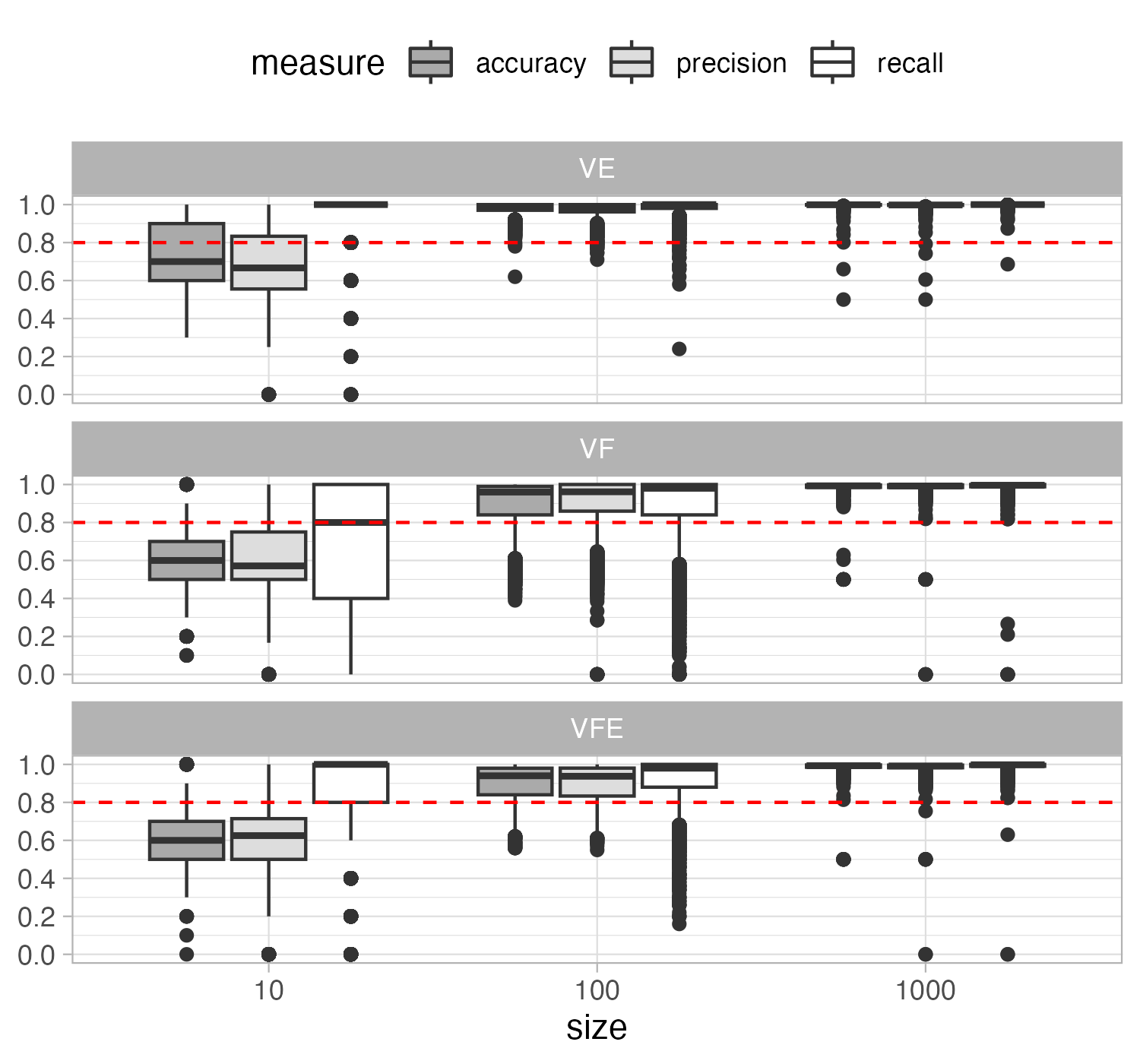}
  \caption{Scores obtained by the classifiers according to the number of lines used for training, grouped by ratio, with the balanced validation}
\label{fig:balanced_corpus_scores}
\end{figure}

\Cref{fig:balanced_corpus_scores} depicts the results obtained with the balanced validation.
First, regardless of size and ratio, the full set of 34,200 classifiers (38 rules $\times$ 3 sizes $\times$ 3 ratios $\times$ 100 measures) obtains a median precision of $0.979$ and a median recall of $1$.
These first results are very promising since we aim for a minimum precision of 0.8 and an ideal precision $\geq 0.95$.

\subsubsection{Influence of size}
\label{subsubsec:results/balanced/size}
Looking at the distributions, we notice that all scores increase as the size of the training set grows, regardless of the ratio.
First, we group all the measures obtained for each size and compare their medians.
Regarding precision, the sizes S, M, and L obtain, respectively, $0.625$, $0.977$, and $0.995$.
Accuracy follows the same trend with $0.700$, $0.970$, and $0.996$.
The recall is stable whatever the size with scores of $1$, $0.980$, and $0.998$.
We observe the same trend on every individual ratio, as shown in~\Cref{fig:balanced_corpus_scores}.

To better qualify the difference between each size, we then compute non-parametric Mann-Whitney U tests~\cite{mann_test_1947} and effect sizes (using rank-biserial correlation---RBC~\cite{kerby_simple_2014}) between the groups (\Cref{tab:pval_rb_training-like}).
The RBC is a value between $-1$ and $1$.
An RBC value of $1$ indicates that all values from the first group are greater than all values from the second group.
A value of 0 indicates an equal amount of values in each group greater than in the other group.
We compare the groups pairwise: S vs.\ M, S vs.\ L, and M vs.\ L.
We adjust the resulting p-values using a Bonferroni correction.
We obtain $p$-values equal to $0$ for the three comparisons, rejecting the null hypotheses.
Regarding effect size, the RBC indicates that $S < M < L$.
The difference between S and M, as well as between S and L, is large with an RBC $\leq -0.7$.
The difference between M and L, on the other hand, is less marked, with an RBC of $-0.34$.

\begin{table}
	\centering
	\caption{Mann-Whitney $p$-values and rank-biserial correlation scores obtained for the analysis of precision in the balanced validation}
	\begin{tabular}{@{}llrr@{}}
        \toprule
        \multicolumn{4}{c}{Size}\\
        \midrule
            Size 1 & Size 2 & p-value & RBC \\
        \midrule
            S      & M      & $0$     & $-0.70$ \\
            S      & L      & $0$     & $-0.76$  \\
            M      & L      & $0$     & $-0.34$ \\
        \midrule
        \multicolumn{4}{c}{Ratio}\\
        \midrule
            Ratio 1 & Ratio 2 & p-value       & RBC \\
        \midrule
            VE      & VFE     & $1.8e^{-301}$ & $0.28$ \\
            VE      & VF      & $3.2e^{-112}$ & $0.17$  \\
            VFE     & VF      & $2.5e^{-38}$  & $-0.10$ \\
        \bottomrule
    \end{tabular}
	\label{tab:pval_rb_training-like}
\end{table}

As explained in \Cref{subsec:coding_pratices_and_linters}, developers favor tools that keep false positives to a minimum.
Thus, we now look at how many of the 38 rules produce classifiers that obtain a median precision above our minimum goal of $0.8$ and our ideal goal of $0.95$ (\Cref{tab:number_rules_upper_goal}).
We observe that every rule reaches the minimum and ideal goals for size L, that some do not reach the goals for size M, and that size S cannot produce satisfactory classifiers.

\begin{table}
	\centering
	\caption{Number of rules where the median precision of classifiers is greater than $P$ for each size and ratio in the balanced validation}
	\begin{tabular}{@{}lrrrrrr@{}}
        \toprule
	        & \multicolumn{3}{c}{$P = 0.8$} & \multicolumn{3}{c}{$P = 0.95$} \\
        \cmidrule(r){2-4} \cmidrule(l){5-7}
            Size & VE & VFE & VF & VE & VFE & VF \\
        \midrule
            S & $1$  & $0$   & $3$ & $0$  & $0$   & $1$\\
            M & $38$ & $32$  & $30$ & $38$ & $17$  & $25$\\
            L & $38$ & $38$  & $38$ & $38$ & $38$  & $38$ \\
        \bottomrule
    \end{tabular}
    \label{tab:number_rules_upper_goal}
\end{table}

In summary, we observe that the number of lines used to learn positively affects the performance of the resulting classifiers.
A crucial element is that even for a medium size of 100 lines, we obtain many results that reach our requirements and that these results are close to those obtained with the large size of 1,000 lines.

\subsubsection{Influence of ratio}
\label{subsubsec:results/balanced/ratio}
We apply the same methodology we used to study the influence of size to study the influence of the ratio, and we immediately see that the impact is less apparent.

We analyze the median measures by grouping them by ratio.
For the VE, VFE, and VF ratios, we get precision medians of $0.992$, $0.943$, and $0.978$, accuracy medians of $0.990$, $0.940$, and $0.960$, and recall medians of $1$, $0.998$, and $0.988$.
We observe that the VE ratio performs best, followed by VF and VFE.
When grouping the measures by size (\Cref{fig:balanced_corpus_scores}), we observe that VE performs best for sizes S and M, the results for size L being very similar for the three ratios.
There is no visible difference between the VFE and VF ratios for sizes S and L.

The statistical tests confirm our first observation: the difference between the three configurations is not as marked as with size (\Cref{tab:pval_rb_training-like}).
The three pairwise comparisons are now:~VE vs.\ VFE, VE vs.\ VF, and VFE vs.\ VF.
The results obtained by the classifiers with different ratios are different, as the $p$-values indicate.
However, the RBC scores indicate that their impact is much weaker than the size.

Computing the number of rules with a median precision higher than our thresholds does not help to discriminate the ratios.
For size S, VF appears to perform better; for size M, VE performs better; and for size L, all rules reach the ideal goal of $0.95$.

In summary, the ratio of non-compliant and compliant code (fixed or extant) used to train our classifiers has a limited effect:~the best ratio across all training set sizes is VE, but the improvement is minor.
For this balanced validation, we conclude that the best-performing learning configuration has a size L and a ratio VE.
However, the results obtained with size M also reach our goals, making it usable and easier to apply in our application domain.

\subsection{Realistic Validation}
\label{subsec:results/real_world}

\begin{figure}
  \includegraphics[width=\linewidth]{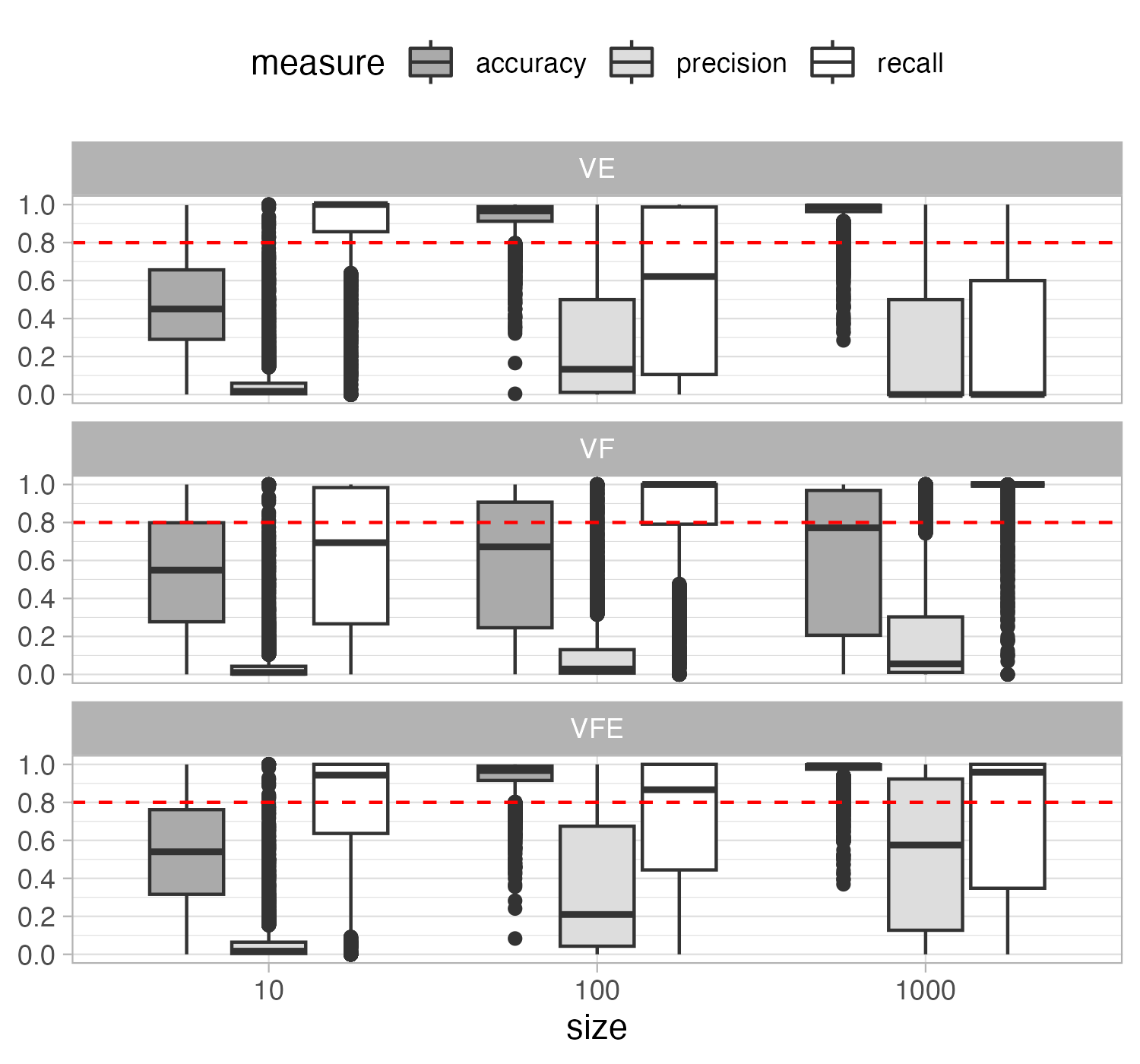}
  \caption{Scores obtained by the classifiers according to the number of lines used for training, grouped by ratio, with the realistic validation}
  \label{fig:ground_truth_with_violation_scores}
\end{figure}

\Cref{fig:ground_truth_with_violation_scores} depicts the results obtained with the realistic validation.
The first observation is clear:~the precision scores plummet, regardless of the size and ratio used for training.
We obtain good results for the global median of the accuracy ($0.887$) and recall ($0.929$), but the median precision falls to $0.043$, compared to $0.979$ with the balanced validation.
 
\subsubsection{Influence of size}
\label{subsubsec:results/real_world/size}
Regarding size, we observe the same tendency as with the balanced validation:~larger sizes for the training set yield better results.
For the sake of conciseness, we do not run a statistical analysis as detailed as for the balanced validation because the obtained precision scores are low across the board:~$0.014$ for size S, $0.091$ for size M, and $0.133$ for size L.
Yet, the statistical tests confirm two points (\Cref{tab:pval_rb_realistic}).
First, there is indeed a difference between the distributions of precisions obtained for each size, and the rank-biserial correlations indicate that $S < M$ and $S < L$.
Second, the RBC score obtained between M and L is only $-0.02$, indicating a very small effect.
Although the precision scores are much lower than observed in the balanced validation, the trend is similar.

\subsubsection{Influence of ratio}
\label{subsubsec:results/real_world/ratio}
With the balanced validation, no ratio configuration stood out.
Here, one configuration produces slightly better results.
Between VF and VE, tests indicate two similar populations, but they are both dominated by VFE.
In our realistic validation, the ratio  has an actual impact on the resulting precision of the classifiers.

\begin{table}
	\centering
	\caption{Mann-Whitney $p$-values and rank-biserial correlation scores obtained for the analysis of precision in the realistic validation}
	\begin{tabular}{@{}llrr@{}}
        
        \toprule
        \multicolumn{4}{c}{Size}\\
        \midrule
            Size 1 & Size 2 & p-value  & RBC \\
        \midrule
            S      & M      & $0$      & $-0.39$ \\
            S      & L      & $0$      & $-0.30$  \\
            M      & L      & $0.0005$ & $-0.02$ \\
        \midrule
        \multicolumn{4}{c}{Ratio}\\
        \midrule
            Ratio 1 & Ratio 2 & p-value       & RBC \\
        \midrule
            VE      & VFE     & $5.0e^{-219}$ & $-0.2408$ \\
            VE      & VF      & $0.31$        & $0.0006$  \\
            VFE     & VF      & $0$           & $0.2909$ \\
        \bottomrule
    \end{tabular}
	\label{tab:pval_rb_realistic}
\end{table}

\subsection{Discussion}
\label{subsubsec:results/discussion}

As a general conclusion of our analyses, we observe that the validation method has a staggering impact on the resulting precision of the classifiers.
Indeed, many rules and classifiers that obtained good precision results in the balanced validation fell off in the realistic validation.
As our industrial scenario involves the analysis of real-world files, where the ratio of non-compliant code over compliant code is generally very low, the classifiers would not perform well w.r.t. to the developers' requirements.
We may explain this lack of success with two points.

First, although CodeBERT fits well with transfer learning, it is usually used to learn on training sets larger than what we require for our scenario.
Our goal is to use as few examples as possible to make the approach usable in practice, so we focus on training set sizes of up to 1,000 lines of code.
It is likely that we would obtain much better results using bigger training set sizes, as was already observed by other authors~\cite{ochodek_recognizing_2020}, but this would hurt the applicability of the approach to our target scenario.
In future work, we will investigate how the approach of~\cite{ochodek_recognizing_2020} behaves with smaller training set sizes, and compare its validation method (stratified k-fold) against ours.

Second, the realistic validation obtains low precision results because there are only a few lines of non-compliant code in the validation set given to the classifiers.
As a reminder, 35 of the 38 rules have a ratio of non-compliant lines over compliant lines lower than 1\%.
The classifiers are battling an extremely imbalanced data problem:~even if the classifiers reach very high accuracy, they will inevitably misclassify a small portion of the compliant lines as non-compliant, which hurts the resulting precision when compliant lines are the most frequent case, due to the base rate fallacy.

In summary, we answer our research questions as follows:

\textbf{RQ1:} With both validation methods, we observe that bigger training set sizes yield better results.
A small size S of 10 lines is clearly insufficient to efficiently learn a coding practice.
On the other hand, in the balanced validation, we observe that the difference between sizes M and L is small and that both can be used to efficiently learn practices with a precision $\geq 0.8$.
Although size M is able to learn some coding practices with a precision $\geq 0.95$, size L performs better in this case.

\textbf{RQ2:} In the balanced validation, we do not observe that a ratio has a clear advantage over the other ones.
In the realistic validation, the VFE ratio obtains better scores, but the precision reached in every case is too low regardless.

\subsection{Threats to Validity}
\label{subsec:results/ttv}
Regarding internal validity, we drew popular projects from GitHub and analyzed all the JavaScript code they contain.
However, some of this code, such as obfuscated code, minified code, or generated code, is not meant to be subjected to coding practices and therefore is not a good basis to learn from and to test against.
We mitigated this threat by doing our best to exclude minified code from our dataset, but we cannot guarantee that our dataset only contains code that has been manually written by developers.

Regarding external validity, our study uses only a subset of rules from ESLint that can be decided on one line.
Therefore, we cannot guarantee that the results obtained on an arbitrary set of company-specific rules would be similar.
However, we made sure to select rules that are possible to learn.
Therefore, our results could be seen as a sort of upper bound.
\section{Related work}
\label{sec:rw}

To the best of our knowledge, the work of Ochodek et al.~\cite{ochodek_recognizing_2020} is the only one that uses machine learning to learn coding practices.
In contrast to our work, they rely on a custom-made feature extractor and classical decision trees for classification.
They rely on a repeated stratified k-fold validation approach to evaluate the accuracy of their classifiers.
They report good F-scores and accuracy on most rules and obtain variable precision scores, which tend to be higher than the ones we obtain on training sets of comparable size (1000 in our study, around 2000 in their study).
However, we push the experiment further by studying the performance of classifiers trained on even smaller training sets.
In the remainder of the section, we discuss other related work on similar topics.

\subsection{Patch Inference in Software Engineering}

Patch inference has been used to automatically repair warnings raised by linters~\cite{markovtsev_style-analyzer_2019,bader_getafix_2019,marcilio_spongebugs_2020,liu_mining_2021,loriot_styler_2022} and correct the use of deprecated APIs, with good reported results~\cite{andersen_generic_2010,meng_lase_2013,jiang_inferring_2019,bavishi_phoenix_2019,serrano_spinfer_2020,haryono_characterization_2021,haryono_androevolve_2022}.
The idea is to automatically infer an AST-based tree transformation from a set of examples, by first using an AST differencing algorithm (such as GumTree~\cite{falleri_fine-grained_2014}), and then abstracting some elements from the produced edit-script.
Some approaches go even beyond and can deal with control and data flow dependencies~\cite{serrano_spinfer_2020}.
In contrast to our approach, they focus on fixing the warnings while we aim at issuing the warnings, with the work of Garg et al.~\cite{garg_synthesizing_2022} being a notable exception since it has the same objective as we do.
An advantage of these approaches is that they may produce human-readable patches that can be manually improved to improve accuracy if needed.
Another advantage is that they appear to require fewer examples than traditional machine learning approaches.
However, since they operate at the AST level, they require advanced language-specific tooling.
Therefore, adapting these approaches to a wide range of languages requires significant effort.
In contrast, our approach uses binary classification and transfer learning which is more straightforward to port to new languages, as it only requires the existence of a pre-trained model for the chosen language.

\subsection{Machine learning for smell detection}

Several approaches use machine learning to detect code smells. A comprehensive overview of these approaches can be found in a survey on the subject~\cite{azeem_machine_2019}.
The main difference between coding practices and code smells is that code smells are generally defined at the design level, while coding practices are defined at the code level.
Therefore the features used to detect code smells are commonly design metrics (such as coupling or cohesion).
Interestingly, none of the surveyed approaches uses a code embedding technique such as CodeBERT.
To the best of our knowledge, only one approach~\cite{kovacevic_automatic_2022} uses code embedding techniques (namely code2vec, code2seq, and CuBERT) in addition to metrics to detect code smells (long method and god class).
In the reported results, CuBERT outperformed all other tested combinations.
This result is consistent with our intuition that BERT-based approaches are well suited to tackle this kind of learning task.

\section{Conclusion}
\label{sec:conclusion}

In this article, we presented a feasibility study on the ability of CodeBERT to learn coding practices through transfer learning using few examples.
While our approach obtains fairly good accuracy and recall, its precision is too low to be usable, probably due to the imbalance between compliant and non-compliant code in real-world software.
An interesting finding is that we obtain weaker precision results than a previous approach~\cite{ochodek_recognizing_2020} which uses a custom feature extraction approach and classical decision trees on training sets of comparable sizes.
This was surprising as CodeBERT generally outperforms classical classifiers in software engineering tasks.
As future work, we will perform a full-fledged replication study of~\cite{ochodek_recognizing_2020} to better understand the influence of the underlying technology on the performance of the classifiers and explore new approaches to classification, such as anomaly detection models.

\bibliographystyle{IEEEtran}
\balance
\bibliography{IEEEabrv,bibfile}

\end{document}